\documentclass[intlimits,twoside,a4paper]{article}

\usepackage{amsmath,amssymb}
\usepackage{graphicx}
\usepackage{wrapfig}
\usepackage{siunitx}
\usepackage{cite}

\usepackage[T2A]{fontenc}
\usepackage[cp1251]{inputenc}

\usepackage[eqsecnum]{cmpj2}

\issue{2013}{16}{1}{13704}

\doinumber{10.5488/CMP.16.13704}

\title[LPRG for Ising-like systems]%
{Linear perturbation renormalization group method for Ising-like spin systems}
\author[J. Sznajd] {J. Sznajd}
\address{Institute for Low Temperature and Structure Research, Polish Academy of Sciences, Wroclaw
}

\date{Received July 4, 2012, in final form January 3, 2013}

\begin{document}

\maketitle

\begin{abstract}
The linear perturbation group transformation (LPRG) is used to study the thermodynamics of the axial next-nearest-neighbor Ising model with four spin interactions (extended ANNNI) in a field.  The LPRG for weakly interacting  Ising chains is presented. The method is used to study finite field para-ferrimagnetic phase transitions observed in  layered uranium compounds, UAs$_{1-x}$Se$_x$, UPd$_2$Si$_2$ or UNi$_2$Si$_2$. The above-mentioned systems are made of ferromagnetic layers and the spins from the nearest-neighbor and next-nearest-neighbor layers are coupled by the antiferromagnetic interactions $J_1 < 0$ and $J_2 < 0$, respectively. Each of these systems exhibits a triple point in which two ordered phases (ferrimagnetic and incommensurate) meet the paramagnetic one, and all undergo the high field phase transition from para- to ferrimagnetic ($++-$) phase. However, if in UAs$_{1-x}$Se$_x$ the para-ferri phase transition is of the first order as expected from the symmetry reason,  in UT$_2$Si$_2$ (T${}={}$Pd, Ni)  this transition seems to be a continuous one, at least in the vicinity of the multicritical point. Within the MFA, the critical character of the finite field para-ferrimagnetic transition at least at one isolated point can be described by the ANNNI model supplemented by an additional, e.g., four-spin interaction. However, in LPRG approximation for the ratio $\kappa = J_2/J_1$ around $0.5$ there is  a critical value of the field for which an isolated critical point also exists in the original ANNNI model.  The positive four-spin interaction shifts the critical point towards higher fields and changes the shape of the specific heat curve. In the latter case for the fields small enough, the specific heat exhibits two-peak structure in the paramagnetic phase.
\pacs 75.10.Hk, 75.40.Cx
\end{abstract}

\section{Introduction}

The Linear Perturbation Renormalization Group (LPRG) \cite{JS} method uses a simple one-dimensional decimation to study universal (critical) and non-universal such as a location of the critical temperature and temperature or field dependence of the properties of thermodynamic quantities of several classical and quantum higher-dimensional models. For the first time, this kind of method was proposed by Suzuki and Takano (ST) \cite{ST}. In the ST approach the one-dimensional decimation is combined with the Migdal-Kadanoff (MK) bond moving approximation. The disadvantages of the latter method especially for the quantum systems were discussed by Barma et al. \cite{Barma} and Castellani et al. \cite{Cas}. Here, we wish only to remind that the MK approach gives rather poor quantitative results even for the two-dimensional Ising model and there is no possibility to construct any systematic approximation procedure within this method. For example, the MK procedure gives for the Ising model on the square lattice the values of the inverse critical temperature $k_{\mathrm{c}} \approx 0.61$ whereas the exact value is  $k_{\mathrm{c}} \approx 0.44$ and for the $s=\frac{1}{2}$ XY model  $k_{\mathrm{c}} \approx 1.2$ \cite{ST} much larger than the value $k_{\mathrm{c}} \approx 0.64$ estimated from the high-temperature series expansion \cite{Jos}, $k_{\mathrm{c}} \approx 0.71(67)$ found from Monte Carlo simulations by fitting to the exponential law \cite{DM,DM_2} and power law \cite{Jonssos}, respectively or rotationally invariant non-linear (block) transformation $k_{\mathrm{c}} \approx 0.62$ \cite{JS1,JS1_2}. The LPRG method has been proposed for the so-called quasi-one-dimensional magnets made of spin chains with the intrachain coupling $k$ and much weaker interchain coupling $k_1 <k$. However, even for the standard Ising model  $k_1=k$, the LPRG approximation gives the results which are in very good agreement with the exact ones $k_{\mathrm{c}} \approx 0.45$ . For $k > k_1 > 0.15 k$, the deviation from the critical temperature exact values is less than $2 \%$ and for $0.5 k > k_1 > 0.15 k$ even less than $1 \%$ \cite{JS2}.

In this paper, the LPRG is used to study the thermodynamics and the existence of a critical point in two-dimensional axial next-nearest-neighbour Ising model with four spin interactions (extended ANNNI) in a field.

\section{LPRG}

  We first describe in detail the LPRG approach in the simplest possible case, i.e., the Ising chains with intrachain interaction $J$ coupled by the weak interchain interactions $J_1$ defined by the Hamiltonian
\begin{equation}
\label{eq:2.1}
 {\mathcal{H}} = k\sum_{\langle ij \rangle}
 S_{i,j}S_{i,j+1} +k_{1}
 \sum_{\langle ij \rangle}S_{i,j}S_{i+1,j}\,,
 \end{equation}
where the label $i$ refers to rows and $j$ refers to columns, the factor $-1/k_{\mathrm{B}}T$ has already been absorbed in the Hamiltonian ($k \equiv J/k_{\mathrm{B}}T$, $k_1 \equiv J_1/k_{\mathrm{B}}T$), and $k_1 < k$.
The LPRG approach starts with an exact decimation for one-dimensional system. Dividing the spins of the chain into two groups, in the simplest case, $s_{2i+1}$ (odd spins~--- survived) and  $s_{2i+2}$ (even spin~--- decimated) one can write the decimation transformation in the form
\begin{equation}
\label{eq:2.2}
\re^{{\mathcal{H}}(s_{2i+1})} = \mathrm{Tr}_{s_{2i+2}}\re^{{\mathcal{H}}(s_{2i+1},s_{2i+2})}.
\end{equation}
In each step of the transformation (\ref{eq:2.2}), every other spin is decimated.
The same transformation can be written by using a linear weight operator $P(\sigma_i, s_i)$
\begin{equation}
\label{eq:2.3}
\re^{{\mathcal{H'}}(\sigma)} = \mathrm{Tr}_{s} P(\sigma,s) \re^{{\mathcal{H}}(s)},
\end{equation}
where the weight operator $P(\sigma,s)$ which couples the original $(s)$ to the new spins $(\sigma)$ is chosen in a linear form as follows:
\begin{equation}
\label{eq:2.4}
P(\sigma,s) = \frac{1}{2N}\prod_{i=1}^N \left(1+\sigma_{i+1}s_{2i+1}\right).
\end{equation}

The weight operator  $P(\sigma,s)$ projects the original spin $s$ space onto the space of the effective spins $\sigma$.
For the one dimensional Ising model with interchain interaction $k_1 = 0$, the transformation (\ref{eq:2.2}) or equivalently (\ref{eq:2.3}) can be carried out exactly. So, we can separate the Hamiltonian (\ref{eq:2.1}) in a manageably exactly unperturbated part ${\mathcal H}_0$ containing intrachain interaction $k$ [first term of the Hamiltonian (\ref{eq:2.1})],  and a remainder ${\mathcal H}_I$ containing interchain interaction $k_1$ [second term in equation~(\ref{eq:2.1})].
With the notation
\begin{equation}
\label{eq:2.5}
z_0 = {\mathrm{Tr}}_s P(\sigma,s) \re^{{\mathcal{H}_0}(s)}
\end{equation}
and
\begin{equation}
\label{eq:2.6}
\langle A\rangle_0 = \frac{1}{z_0}{\mathrm{Tr}}_s A P(\sigma,s) \re^{{\mathcal{H}_0}(s)}
\end{equation}
the transformation (\ref{eq:2.3}) can be written as
\begin{equation}
\label{eq:2.7}
{\mathcal H'}(\sigma) = \ln z_0 + \ln \langle \re^{{\mathcal{H}_I}(s)}\rangle,
\end{equation}
with the following cumulant expansion  for $\ln \langle \re^{{\mathcal{H}_I}(s)}\rangle$  \cite{NvL}
\begin{equation}
\label{eq:2.8}
\ln \langle \re^{{\mathcal{H}_I}(s)}\rangle = \langle{\mathcal{H}}_I\rangle_0 + \frac{1}{2!} \left(\langle {\mathcal{H}_I}^2 \rangle_0 - \langle{\mathcal{H}_I}\rangle^2_0\right).
\end{equation}

\begin{figure}[!t]
\centerline{\includegraphics[width=0.5\textwidth]{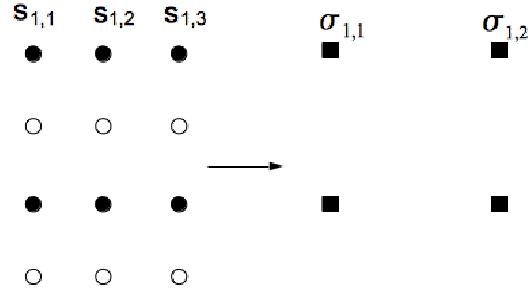}}
\caption{The LPRG procedure for weakly interacting chains in two dimensions.} \label{fig1}
\end{figure}

The idea of LPRG in two dimensions is presented in figure~\ref{fig1}. The chains are divided into two black and white groups. In each renormalization step, every other spin from a black chain is decimated and in a white chain all spins are removed. As a result, one gets the system of effective spins $\sigma$.  The single chain ``partial'' partition function $z_0$ can be easily found as \cite{Nau}
\begin{equation}
\label{eq:2.9}
z_0 = (\cosh 2k +1) +  (\cosh 2k - 1) \sigma_{i,j} \sigma_{i,j+1}\,,
\end{equation}
and
\begin{equation}
\label{eq:2.10}
\ln z_0 = \frac{1}{2} (\ln 2 + \ln \cosh 2k ) +   \frac{1}{2}  \sigma_{i,j} \sigma_{i,j+1 } \ln \cosh 2k .
\end{equation}

To evaluate the cumulants (\ref{eq:2.8}), one has to know the averages $\langle s_{i,j}\rangle$ and $\langle s_{i,j}\dots s_{i,j+n}\rangle$ from the black (decimated) and white (removed) rows. For the spins from the black rows (figure~\ref{fig1})
\begin{equation}
\label{eq:2.11}
\langle s_{1,1}\rangle = \sigma_{1,1}\,, \qquad  \langle s_{1,2}\rangle = \frac{1}{2}(\sigma_{1,1}+\sigma_{1,2}) \tanh 2k , \qquad \langle s_{1,3}\rangle = \sigma_{1,2}\,,
\end{equation}
and
\begin{equation}
\label{eq:2.12}
\langle s_{1,1} s_{1,2}\rangle =\frac{1}{2} (1+\sigma_{1,1} \sigma_{1,2}) \tanh 2k , \qquad  \langle s_{1,1} s_{1,3}\rangle = \sigma_{1,1} \sigma_{1,2}\,.
\end{equation}
For the spins from the white rows
\begin{equation}
\label{eq:2.13}
\langle s_{2,j}\rangle = 0, \qquad \langle s_{2,j} s_{2,j+n}\rangle = \tanh^n k.
\end{equation}
Now it is relatively easy to find the renormalized Hamiltonian ${\mathcal{H'}}(\sigma)$ (2.7) in the form
\begin{equation}
\label{eq:2.14}
{\mathcal{H'}}(\sigma) = G(k_i) + k' \sum \sigma_{i,j} \sigma_{i,j+1} + k'_1 \sum \sigma_{i,j} \sigma_{i+1,j} + \Omega(\sigma),
\end{equation}
where $G(k_i)$ is a constant (independent of effective spins $\sigma$) term which can be used to calculate the free energy per site according to the formula
\begin{equation}
\label{eq:2.15}
f = \sum_{n=1}^{\infty} \frac{G(k_i^{  (n)  })}{3^n}\,,
\end{equation}
and  ``$n$'' numbers the LPRG steps. $\Omega(\sigma)$ denotes the additional interactions generated eventually by LPRG transformation.

The approach presented above can be also used to consider a quantum spin model or an interacting electron model. However, in these cases the LPRG does not start with an exact but with an approximate decimation for one-dimensional systems \cite{ST}. We should also emphasize  that generally, the approach is relevant for higher temperatures. Using LPRG one can show the existence of the critical point, and can obtain the location of a transition point, if any, and the temperature or field dependences of the thermodynamic quantities.

\section{Extended ANNNI model in a field}

There exists a class of compounds with a layered structure and strong c-axial anisotropy which can be described by the Ising-like ANNNI model or its extensions. For example, models of this kind have been proposed to describe the phase diagrams of the uranium compounds:  UAs$_{1-x}$Se$_x$ with $x<0.1$ \cite{RM} or UT$_2$Si$_2$ (T${}={}$Pd, Ni) \cite{NS,honma} made of ferromagnetic layers which comprise three ordered phases: antiferromagnetic $(+-+-)$, ferrimagnetic $(++-)$ and incommensurate. Each of this system exhibits a triple point in which two ordered phases (ferrimagnetic and incommensurate) meet the paramagnetic one, and all undergo the high field phase transition from paramagnetic to ferrimagnetic $(++-)$ phase. However, if in UAs$_{1-x}$Se$_x$ the para-ferri phase transition is of the first order, in UT$_2$Si$_2$ this transition, at least in the vicinity of the multicritical point,  seems to be a continuous one.
For the symmetry reason, the para-ferrimagnetic $(++-)$ phase transition in the presence of an external field should be discontinuous and a question has arised if the ANNNI model can exhibit in such a case a critical transition. Within the MFA it has been shown that the ANNNI model should be supplemented by some additional, e.g., four-spin interaction to exhibit an isolated critical point \cite{PKS}. In order to answer this question, we have used the LPRG to study the extended ANNNI model in a field
\begin{eqnarray}\label{eq:3.1}
 {\mathcal{H}} &=& k\sum_{\langle ij \rangle} S_{i,j}S_{i,j+1} +k_{1} \sum_{\langle ij \rangle}S_{i,j}S_{i+1,j} + k_{2} \sum_{\langle ij \rangle}S_{i,j}S_{i+2,j} \\ \nonumber &&{}+ k_{4} \sum_{\langle ij \rangle}S_{i,j}S_{i+1,j}S_{i,j+1}S_{i+1,j+1} + h \sum_{\langle i\rangle}S_{i}\,.
\end{eqnarray}

In the ground state in zero field, the standard ANNNI model with $k_4=0$, $k_1, k_2<0$ and $\kappa \equiv  k_2/k_1 = 0.5$ exhibits a multicritical point where antiferromagnetic phases $(+-+-)$ and $(++--)$ meet a ferrimagnetic phase with the spins of the ferromagnetic planes ordered in the sequence $(++-)$ ($\langle 2,1\rangle $ phase in the notation of the paper \cite{honma}). The external magnetic field extends the $(++-)$ phase in the $(\kappa, H)$ plane \cite{Uimin}. All magnetic structures of UPd$_2$Si$_2$ as well as UNi$_2$Si$_2$ and UAs$_{1-x}$Se$_x$ ($0 \leqslant x \leqslant 0.1$) in the fields along $c$ axis exhibit ferromagnetic layers with moments perpendicular to the layers. So, only the magnetic order between the ferromagnetic layers is of interest. Consequently, it seems that to understand the main feature of the paramagnetic-ferrimagnetic ($++-$) phase transition one can confine oneself to consider the two dimensional (2D) problem. Thus, in the present paper we use the LPRG method to study the extended ANNNI model in two dimensions.

\begin{figure}[!b]
\centerline{\includegraphics[width=0.65\textwidth]{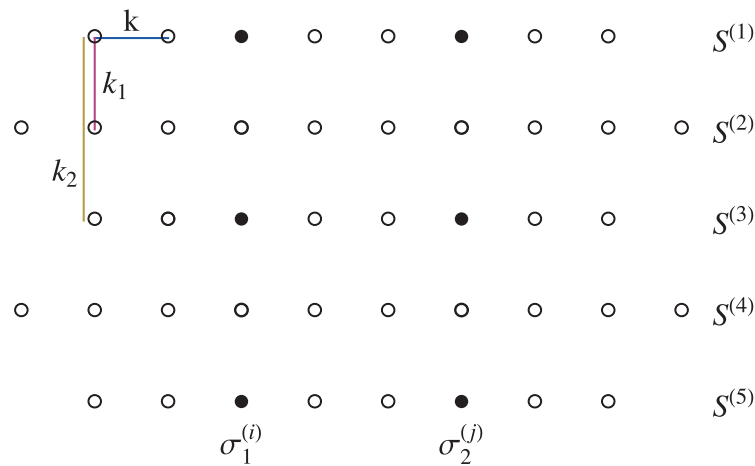}}
\caption{Cluster (8-10-8-10-8) used to get renormalized Hamiltonian of the extended ANNNI model in the LPRG procedure.}
\label{fig2}
\end{figure}

The idea of the LPRG for ANNNI-type model in 2D with the following projector for one decimated row
\begin{equation}
\label{eq:3.2}
P(\sigma,s) = \frac{1}{4} (1+\sigma_{1,1} S_{1,3})(1+\sigma_{1,2} S_{1,6})
\end{equation}
is presented in figure~\ref{fig2}. The full circles represent the spins which survive in the decimation procedure. According to figure~\ref{fig2}, in each step of the RG transformation every other row (``even row'') is removed, and from odd rows every third spin survives. The cluster presented in figure~\ref{fig2} can be used to study systems made of ferromagnetic or antiferromagnetic chains {\emph {but it does not preserve the antiferromagnetic and much less modulated order between the chains}}. However, this cluster preserves the ferrimagnetic ($++-$) ordering  which is a matter of interest in this paper. In order to take into account both possibilities (antiferromagnetic and ferrimagnetic orders) one should consider a cluster of $11$ chains and for three possibilities (antiferromagnetic $+-+-$, antiferromagnetic $++--$ , and ferrimagnetic orders) $16$ chains. Although calculation with such clusters is straightforward it becomes quite involved.
It should be emphasized once more that the LPRG with the cluster presented in figure~\ref{fig2} is suitable to describe, except for para-ferro, only para-ferrimagnetic ($++-$) phase transitions. Such phase transitions can occur in both 2D and 3D systems. Thus, we have confined ourselves to consider 2D problem although the incommensurate phases observed in the mentioned uranium compounds can be described by 3D ANNNI model.

As usual, the RG procedure leads from the set of original parameters of the Hamiltonian (\ref{eq:2.3}) ($h$, $k$, $k_1$, $k_2$, $k_4$) to the set of  renormalized parameters ($h'$, $k'$, $k'_1$, $k'_2$, $k'_4$)
\begin{equation}
\label{eq:3.3}
h' \sigma_j^{i},\qquad k' \sigma_1^{(j)} \sigma_2^{(j)}, \qquad k'_1 \sigma_1^{j} \sigma_1^{(j+1)},\qquad k'_2 \sigma_1^{(j)} \sigma_1^{(j+2)}, \qquad k'_4 \sigma_1^{(j)} \sigma_1^{(j+1)}\sigma_2^{(j)} \sigma_2^{(j+1)}.
\end{equation}
However, in the lowest nontrivial order of the cumulant expansion, which in our case is the second order, several new odd and even interactions come into play. For the sake of simplicity, we will take into account the contributions up to the second order only from the one- and two-spin interactions, contributions to the first order from three- and four-spin interactions, and neglect any contributions from the higher order spin terms  \cite{tbp}. Accordingly, we consider 17 parameters, i.e.,  five original (\ref{eq:3.3}) and 12 generated by the RG transformation.

\begin{figure}[!b]
\centerline{\includegraphics[width=0.6\textwidth]{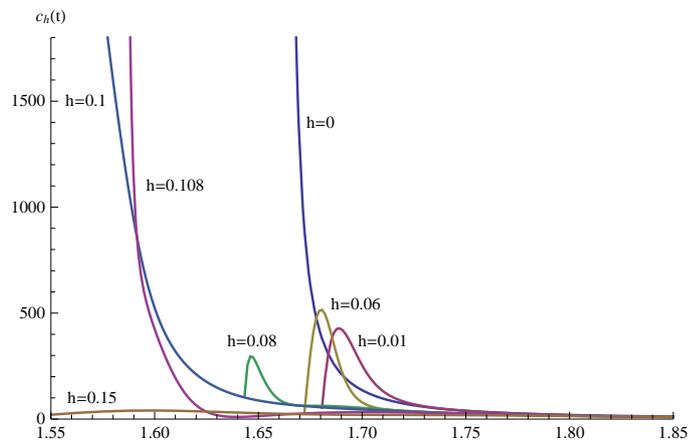}}
\caption{(Color online) The specific heat temperature dependences of the ANNNI model with $k_1=-0.6$, $k_2=-0.3$, $k_4 = 0$ for several values of the field.}\label{fig3}
\end{figure}

To evaluate the transformation (\ref{eq:2.8}) one has to know the averages of spins $\langle S_{i,j}\rangle $ and two spins products $\langle S_{i,j} S_{i,j+n}\rangle $ from the decimated (``odd'') and removed (``even'') rows. The chain averages for the spins from decimated rows of the Ising model in a field with the weight operator (\ref{eq:3.2}) have been presented in paper \cite{JS1}, and for the spins from the removed rows, these averages are known exactly (see for example \cite{Selke}  and references therein). Now, we are able to numerically evaluate the renormalization transformation (\ref{eq:2.8}) which in our approximation has a form of 17 recursion relations.
As usual, in order to determine the critical temperature, one has to find a critical surface which separates in the parameter space the region of attraction of the two stable fixed points, zero temperature, $k_{\alpha} = \infty$ and infinite temperature $k_{\alpha} = 0$. The existence of such a surface means that the system can undergo a second order phase transition.
Moreover, we can also numerically calculate the free energy per spin collecting the constant terms generated in each step of the iteration process (\ref{eq:2.15}). In a disordered phase we obtain a rapidly convergent infinite series for the free energy which can be used to find  the specific heat as a function of the reduced temperature $t = T/J$.

\begin{figure}[!t]
\centerline{\includegraphics[width=0.62\textwidth]{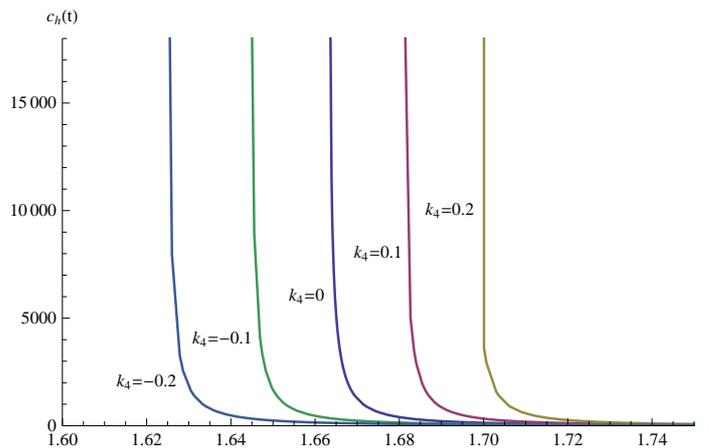}}
\caption{(Color online) The specific heat temperature dependences of the extended ANNNI model with $k_1=-0.6$, $k_2=-0.3$ at zero field for several values of the four spin coupling $k_4$.}
\label{fig4}
\end{figure}
\begin{figure}[!h]
\centerline{\includegraphics[width=0.62\textwidth]{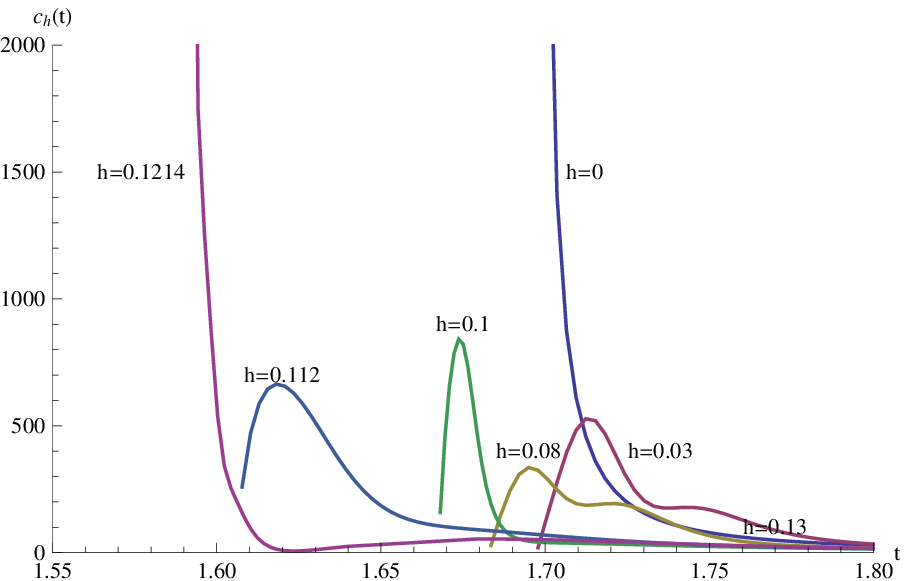}}
\caption{(Color online) The specific heat temperature dependences of the extended ANNNI model with $k_1=-0.6, k_2=-0.3, k_4=0.2$ for several values of the field.}
\label{fig5}
\end{figure}
\begin{figure}[!h]
\centerline{\includegraphics[width=0.63\textwidth]{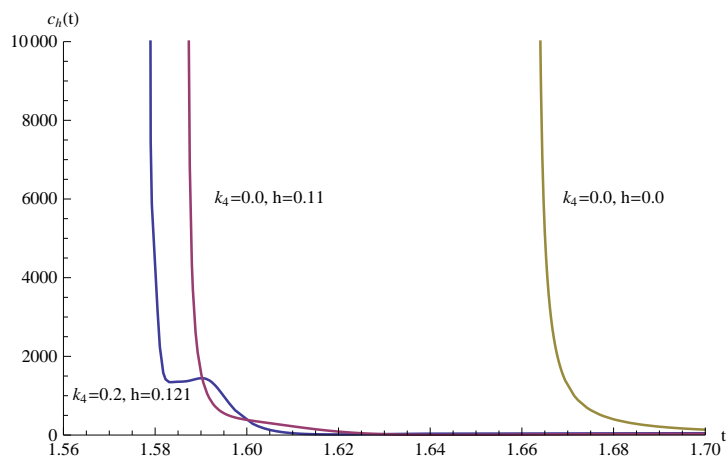}}
\caption{(Color online) Comparison of the specific heat curves of the standard ANNNI model at fields $h=0$ and  $h=0.11$ around the critical field with such a curve for the model with $k_4=0.2$ around the critical field $h=0.121$.}
\label{fig6}
\end{figure}

Let us start with the standard ANNNI model with $k=1, k_1=-0.6, k_2=-0.3$ and $k_4=0$ \cite{tbp}.
The value of the critical temperature has been determined on the ground of the recursion relation analysis. Using the formula (\ref{eq:2.15}), the free energy per site and the specific heat as functions of temperature and a field have been found. In figure~\ref{fig3}, the temperature dependence of the specific heat for several values of the field is presented. The specific heat divergence corresponding to a critical point is visible for $h=0$ and for $h$ around $0.11$. For $0<h<h_{\mathrm{c}} \approx 0.11$, the specific heat exhibits a maximum.  Unfortunately, the LPRG fails to properly  describe the specific heat behavior below the maximum because it is not capable of describing the ordered phase which is expected below the maximum, if the system undergoes a discontinuous phase transition. For some value of the field $h>h_{\mathrm{c}}$  ($h=0.15$ in figure~\ref{fig3}), specific heat has only a broad hump.

This means that the considered ANNNI in the presence of the external field can exhibit a continuous phase transition to the ferrimagnetic phase $(++-)$ at $h=0$ and $h=h_{\mathrm{c}} \approx 0.11$. Thus, within the LPRG unlike in MFA it is not necessary to supplement the original ANNNI model by an additional four-spin interaction to reach an isolated critical point for a finite field around $h = 0.11$ \cite{tbp}.

\begin{figure}[!t]
\centerline{\includegraphics[width=0.65\textwidth]{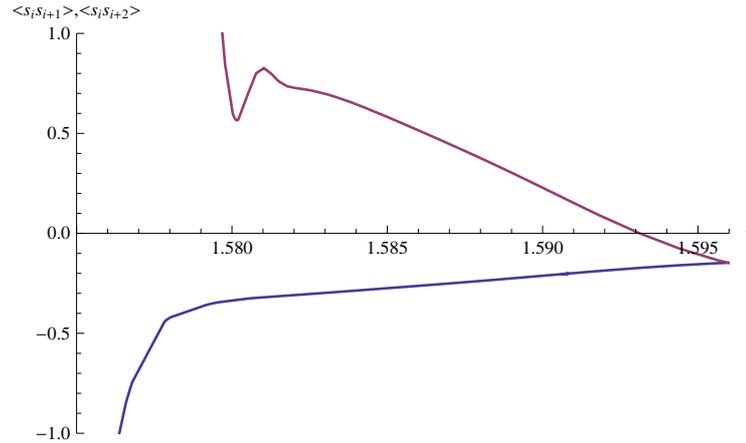}}
\caption{(Color online) The temperature dependences of the nearest-neighbor $\langle s_i s_{i+1}\rangle$ (bottom curve) and next-nearest-neighbor $\langle s_i s_{i+2}\rangle $ chains correlation functions for extended ANNNI $(k_4=0.2)$ model at $h=0.121$.}
\label{fig7}
\end{figure}

Now, we proceed to the extended ANNNI model with $k = 1, k1 = -0.6, k2 = -0.3$ and $k_4 \not = 0$. In figure~\ref{fig4} we present the temperature dependences of the specific heat in zero field for several values of negative and positive values of $k_4$. As seen, the four-spin interaction simply shifts the critical point towards higher temperature for $k_4 >0$ and towards lower temperature for $k_4<0$. Figure~\ref{fig5} shows  the temperature dependence of the specific heat for the model with $k_4=0.2$ in finite fields. Similarly to the standard ANNNI model case, the application of the external field changes the divergence of the specific heat into a maximum for $0 < h < h_{\mathrm{c}}$. For $h \approx 0.1214$, the divergence appears again indicating a critical point. In figure~\ref{fig6} we compare the shapes of the specific heat curves for the field strengths close to the critical values for standard and extended ANNNI models. In the case of the $k_4=0.2$ model, a small peak in specific heat appears  above the critical temperature. In figure~\ref{fig7}, the two-site correlation functions for the spin from adjacent ($G_{1}=\langle s_i s_{i+1}\rangle $) and next nearest neighbor  ($G_2=\langle s_i s_{i+2}\rangle $) chains are presented. As seen, a kink of $G_2$ around the temperature of the specific maximum is observed. It can suggest some preliminary ordering between the spins from the next nearest neighbour chains.
To summarize, it has been shown that within the LPRG approximation, both the original ANNNI model and the one  supplemented by the four-spin interaction with $k_2/k_1 = 0.5$ and $k_1, k_2 < 0$ undergo the continuous phase transition at  $h = 0$ and can undergo such a transition for some finite value of the field $h = h_{\mathrm{c}}$. In the calculations presented in this paper, a sharp anomaly in specific heat is evident at fields around $h_{\mathrm{c}}$ which could explain the experimentally observed anomalies in UPd$_2$Si$_2$. The four-spin interaction shifts both the critical temperature and field, and changes the shape of the specific heat curves.

\ukrainianpart

\title{Метод лінійної пертурбативної ренормалізаційної групи \\ для
ізингоподібних спінових систем}
\author{Й. Шнайд}
\address{Інститут низької температури і структурних досліджень, Польська академія наук,
Вроцлав, Польща}

\makeukrtitle

\begin{abstract}
\tolerance=3000
Лінійне пертурбативне перетворення (LPRG) використовується для
вивчення термодинаміки аксіальної моделі Ізинга з наступними до
найближчих сусідами з чотириспіновою взаємодією (розширена модель
ANNNI) у полі.  Представлено LPRG для слабовзаємодіючих ланцюжків
Ізинга. Метод застосовано до вивчення пара-феромагнітних фазових
переходів у скінченному полі, що спостерігаються в шаруватих
сполуках урану UAs$_{1-x}$Se$_x$, UPd$_2$Si$_2$ чи UNi$_2$Si$_2$.
Вище згадані системи зроблені з феромагнітних шарів і спіни з
найближчих і наступних до найближчих шарів є зв'язані
антиферомагнітними взаємодіями $J_1 < 0$ і $J_2 < 0$, відповідно.
Кожна з цих систем демонструє потрійну точку, в якій дві
впорядковані фази (феромагнітна і неспівмірна) зустрічаються з
парамагнітною фазою і всі фази зазнають  фазового переходу у
сильному полі з пара- до феромагнітної ($++-$) фази. Проте, якщо в
UAs$_{1-x}$Se$_x$ є пара-феро фазовий перехід першого роду, як
очікується з симетрійних міркувань, в UT$_2$Si$_2$ (T${}={}$Pd, Ni)
цей перехід видається неперервним, принаймні в околі мультикритичної
точки. В рамках наближення середнього поля, критич\-ний характер
пара-феромагнітного переходу в скінченному полі, принаймні в одній
ізольованій точці, може бути описаний за допомогою моделі ANNNI, яка
доповнена додатковою, наприклад, чотириспіновою взаємодією. Проте, в
наближенні  LPRG для коефіцієнта  $\kappa = J_2/J_1$ поблизу $0.5$ є
критичне значення поля, для якого ізольована критична точка також
існує в оригінальній моделі ANNNI. Позитивна чотириспінова взаємодія
зсуває критичну точку до вищих полів і змінює форму кривої питомої
теплоємності. В останньому випадку для достатньо малих полів питома
теплоємність демонструє двопікову структуру в парамагнітній фазі.

\keywords модель ANNNI, ренормалізаційна група, ізольована критична
точка
\end{abstract}

\end{document}